\documentclass[%
 reprint,
 amsmath,amssymb,
 aps,
]{revtex4-2}

\usepackage{graphicx}% Include figure files
\usepackage{dcolumn}% Align table columns on decimal point
\usepackage{bm}% bold math
\begin{document}

%\preprint{APS/123-QED}

\title{Neutrino oscillations from the perspective of the quantum Yang-Baxter equations }% Force line breaks with \\
%\thanks{A footnote to the article title}%

%\author{W. Rosado}
 %\altaffiliation[Also at ]{Physics Department, XYZ University.}%Lines break automatically or can be forced with \\
%\author{Second Author}%
%\email{wilson.rosado@unisucre.edu.co}
%\affiliation{%
% Departamento de Física, Universidad de Sucre,\\
% Cra 28 No 5-267 Puerta Roja, Sincelejo, Colombia}%

%\collaboration{MUSO Collaboration}%\noaffiliation

\author{Ivan Arraut}
%\homepage{http://www.Second.institution.edu/~Charlie.Author}
\email{ivan.arraut@usj.edu.mo}
\affiliation{
University of Saint Joseph,\\
Estrada Marginal da Ilha Verde, 14-17, Macao, China
}%
\author{Enrique Arrieta-Diaz}
%\homepage{http://www.Second.institution.edu/~Charlie.Author}
\email{arrieta1@fnal.gov}
\affiliation{Universidad del Magdalena,\\
Carrera 32 No 22 – 08 Santa Marta, Colombia
}
%\affiliation{
% Third institution, the second for Charlie Author
%}%
%\author{Delta Author}
%\affiliation{%
 %Authors' institution and/or address\\
 %This line break forced with \textbackslash\textbackslash
%}%

%\collaboration{CLEO Collaboration}%\noaffiliation

\date{\today}% It is always \today, today,
             %  but any date may be explicitly specified

\begin{abstract}
The origins of neutrino masses is one of the biggest mysteries in modern physics since they are beyond the realm of the Standard Model. As massive particles, neutrinos undergo flavor oscillations throughout their propagation. In this paper we show that when a neutrino oscillates from a flavor state $\alpha$ to a flavor state $\beta$, it follows three possible paths consistent with the Quantum Yang-Baxter Equations. These trajectories define the transition probabilities of the oscillations. Moreover, we define a probability matrix for flavor transitions consistent with the Quantum Yang-Baxter Equations, and estimate the values of the three neutrino mass eigenvalues within the framework of the triangular formulation.
%\begin{description}
%\item[Usage]
%Secondary publications and information retrieval purposes.
%\item[Structure]
%You may use the \texttt{description} environment to structure your abstract;
%use the optional argument of the \verb+\item+ command to give the category of each item. 
%\end{description}
\end{abstract}

%\keywords{Suggested keywords}%Use showkeys class option if keyword
                              %display desired
\maketitle

%\tableofcontents

\section{Introduction}

The neutrino was initially conjectured as a necessity for restoring the energy-momentum conservation in certain processes \cite{the_neutrino}. The electroweak theory, which successfully predicted the existence of the $Z$ and $W$ bosons, was developed under the assumption of massless neutrinos. In principle, only left-handed neutrinos are observed in nature, therefore, massive terms cannot be generated in the Lagrangian of the electroweak theory via Higgs mechanism \cite{5}. However, the Higgs mechanism generates the masses of all other fundamental particles \cite{glashow, salam, weinberg}.  The existence of right-handed neutrinos has not been ruled out experimentally \cite{nova_sterile}, as there is the possibility of the existence of sterile neutrinos \cite{6}, which we will ignore. Nonetheless, the seminal discovery of neutrino oscillations \cite{sk, sno} between their three flavor states \cite{lep, Cosmology}, showed that neutrinos are massive \cite{7}. Hence, neutrino oscillations are parameterized as a superposition of the mass eigenstates producing a flavor state \cite{pontecorvo_2, pontecorvo_3, mns}. In this paper we show that neutrino oscillations are equivalent to oscillations of a particle between three different ground states, without the possibility to select one in particular. Therefore, neutrinos do not spontaneously break a yet-to-be-determined symmetry related to its oscillations. Instead, the particle keeps oscillating by obeying a triangular relation consistent with the Quantum Yang-Baxter Equations (QYBE) \cite{YB}. The triangular formulation, which ignores any possible charge-parity (CP) violation, predicts a normal mass hierarchy with $m_3 > m_2 \approx m_1$. The triangular formulation creates new relations between the mixing angles and the mass eigenvalues, which help to predict the values of $m_1$, $m_2$ and $m_3$, although the mechanism behind the origin of neutrino masses is not yet understood. \\

Here we show that the probability for a neutrino to oscillate from a flavor state $\alpha$ to a flavor state $\beta$, corresponds exactly to the sum of three possible histories of oscillation. These histories can be accommodated inside a triangle obeying the QYBE \cite{Mypaper}. Moreover, we show the full consistency of the triangular formulation with the oscillation probability. Finally, we find probability matrices obeying the QYBE, and interpret them in a probabilistic scenario. This shows the consistency of the patterns of neutrino oscillations with the QYBE, and suggests the existence of a hidden symmetry which is not spontaneously broken.

%propose a possible mechanism for the origin of the neutrino mass oscillation. The mechanism consists on a mixing term which shows the interaction between left-handed neutrinos and right-handed virtual electrons, muons and tauns. This term emerges as an additional contribution in the electroweak theory and it generates the neutrino masses. We propose some signatures that can be detected in the current experiments with DUNE or NOVA, in order to test this idea.

\section{The neutrino mixing angles}   \label{mix}

The neutrino flavor eigenstates $\nu_\alpha$ are linear combinations of the mass eigenstates $\nu_i$

\begin{equation}   \label{omne}
\nu_\alpha=\sum_iU_{\alpha i}\nu_i,
\end{equation}
where the $U_{\alpha i}$ are the elements of a mixing matrix $U$ \cite{pontecorvo,mns}. In eq. (\ref{omne}), index $\alpha$ corresponds to the flavor eigenstates and index $i$ corresponds to the mass eigenstates. The matrix $U$ is invertible and unitary. For the full three-flavor oscillation case, the mixing matrix is defined as 

\begin{equation}   \label{matrix}
\begin{bmatrix}
c_{12}c_{13} & s_{12}c_{13} & s_{13}\\
-s_{12}c_{23}-c_{12}s_{23}s_{13} & c_{12}c_{23}-s_{12}s_{23}s_{13} & s_{23}c_{13}\\
s_{12}s_{23}-c_{12}c_{23}s_{13} & -c_{12}s_{23}-s_{12}c_{23}s_{13} & c_{23}c_{13}
\end{bmatrix}.
\end{equation}
Here $c_{ij} = cos \left( \theta_{ij} \right)$ and $s_{ij} = sin \left( \theta_{ij} \right)$. The matrix represents a three-dimensional rotation via the Tait-Bryan angles \cite{TB}. Matrix (\ref{matrix}) contains three mixing angles, or parameters, which have been measured by oscillation experiments \cite{nova_theta, kamland, dayabay}. However, no CP violation is considered in matrix (\ref{matrix}), which normally appears as an additional measurable parameter \cite{pontecorvo, mns}. Other parameters that appear in the oscillation probabilities are the differences between the squared
mass eigenvalues, namely, $\Delta m_{21}^2$ and $\Delta m_{32}^2 \approx \Delta m_{31}^2$, which have been measured by oscillation experiments \cite{sk, sno, sk2, nova_cp}. 
\section{Neutrinos and the vacuum states}   \label{Vacuumpers}

The neutrino could be considered as a particle that perceives three different vacuum states when it propagates through spacetime. Hence, we introduce three different ground states corresponding to the electron, muon, and tau neutrinos. Therefore, the vacuum state perceived by the neutrino is a linear superposition of the three ground states $\vert0 \rangle_e$, $\vert0 \rangle_\mu$, and $\vert0 \rangle_\tau$. This is a consequence of the time dependence of the neutrino states and their relationship with the inverse of matrix (\ref{matrix}). We can define the vacuum states as follows

\begin{equation}
\hat{a}^+_e\vert0_e \rangle = \vert\nu_e \rangle,\;\;\;\hat{a}_e\vert0_e \rangle = 0.    
\end{equation}
A neutrino with a mass $m_i$, represents an oscillatory vacuum state, which can be defined as

\begin{equation}   \label{standardvac}
\hat{a}_i\vert0 \rangle_i = 0.    
\end{equation}
The vacuum corresponding to the mass eigenstates is related to the vacuum corresponding to the flavor eigenstates as follows

\begin{equation}   \label{super}
\vert0 \rangle_i=\sum_\alpha a_\alpha \vert0 \rangle_\alpha.
\end{equation}
The coefficients $a_\alpha$ represent the factors containing the mixing angles $\theta_{ij}$. They can be derived directly from the inverse of the matrix (\ref{matrix}). Inverting eq. (\ref{super}) results in 

\begin{equation}
\vert0 \rangle_\alpha=\sum_i b_i \vert0 \rangle_i,
\end{equation}
where the $b_i$ represent other factors containing the mixing angles $\theta_{ij}$. From matrix (\ref{matrix}) we get

\begin{eqnarray} \label{linear_comb}
\vert0 \rangle_e & = & c_{12}c_{13}\vert0\rangle\rangle_1+s_{12}c_{13}\vert0\rangle\rangle_2+s_{13}\vert0\rangle\rangle_3, \nonumber \\
\vert0 \rangle_\mu & = & (-s_{12}c_{23}-c_{12}s_{23}s_{13})\vert0\rangle_1 + \nonumber \\
& & (c_{12}c_{23}-s_{12}s_{23}s_{13})\vert0\rangle_2+s_{23}c_{13}\vert0\rangle_3, \nonumber \\
\vert0 \rangle_\tau & = & (s_{12}s_{23}-c_{12}c_{23}s_{13})\vert0\rangle_1 + \nonumber \\
& & (-c_{12}s_{23}-s_{12}c_{23}s_{13})\vert0\rangle_2+c_{23}c_{13}\vert0\rangle_3.
\end{eqnarray}
Eqns. (\ref{linear_comb}) expand the vacuum for the flavor states as a linear combination of the vacuum corresponding to the mass eigenstates by using the same relations between the effective mass of the flavor states and the mass eigenvalues. The additional relations, connecting the mass eigenvalues with the mixing angles, emerge via the triangular formulation.
 
\section{Neutrino oscillation as a triangular formulation}   \label{YANG}

The triangular formulation of the neutrino mass problem states that the inverse of each mass eigenvalue corresponds to a side of a right triangle. Additionally, the mixing angles multiplied by two, correspond to the internal angles of the triangle. Under the assumption of no CP violation, the triangle satisfies the QYBE. This assumption is favored by \cite{neutrino2024} in the normal mass hierarchy. Under the triangular formulation, the symmetry of the ground state is not spontaneously broken. Hence, the neutrino mass cannot be generated via the Higgs mechanism. \\
\begin{figure}[h!]
\centering
\includegraphics[scale=0.42]{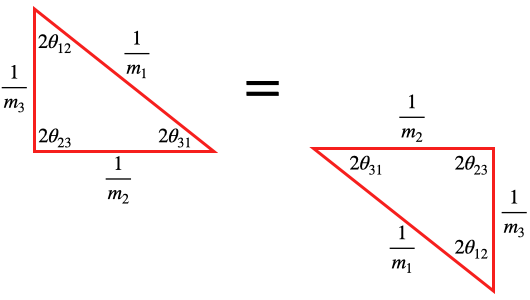} 
\caption{The triangular formulation for neutrino oscillations. The lengths of the sides are matched to the inverses of the mass eigenvalues. The internal angles are matched to the mixing angles times 2. \label{fig1}}
\end{figure}
Our triangular relations \cite{Mypaper}, which constrain the dynamics of neutrinos, are shown in figure (\ref{fig1}). Each side of the triangle corresponds to a value $1/m_i$, with $i=1, 2, 3$. We assume that the triangle is approximately a right triangle, thus $\theta_{23} \approx \pi/2$. This is consistent with the results for $\theta_{23}$ presented in \cite{neutrino2024}. With the Pythagoras theorem we relate the three sides of the triangle

\begin{equation} \label{pitagoras}
\frac{1}{m_1^2}=\frac{1}{m_2^2}+\frac{1}{m_3^2}.
\end{equation}
Combining eq. (\ref{pitagoras}) with the results from \cite{dayabay} on $\theta_{31}$, we obtain the mass ratios

\begin{eqnarray} \label{cosine}
\cos(2\theta_{31}) \approx & \frac{m_1}{m_2} = 0.957, \nonumber \\
\sin(2\theta_{31}) \approx & \frac{m_1}{m_3} = 0.290.
\end{eqnarray} 

These mass ratios are consistent with the differences between the squared mass eigenvalues presented in \cite{Exp}

\begin{eqnarray}   \label{masa1}
m_2^2-m_1^2 \approx 7.41\times10^{-5}eV^2, \nonumber \\    
m_3^2-m_1^2 \approx m_3^2-m_2^2 \approx 2.43\times10^{-3}eV^2.    
\end{eqnarray}
Therefore, we express the mixing angles as a function of the mass eigenvalues by using standard trigonometry.
%
%\begin{equation}   \label{mici}
%sin 2\theta_{13}\approx\frac{m_1}{m_3}, \;\;\;sin 2\theta_{12}\approx\frac{m_1}{m_2},\;\;\;sin 2\theta_{23}\approx 1.    
%\end{equation}
%
Additionally, we have the relation 

\begin{equation}
\theta_{12}+\theta_{13}+\theta_{23}\approx\frac{\pi}{2},    
\end{equation}
which constraints the values of the mixing angles. Combining eqns. (\ref{cosine}) and (\ref{masa1}) yield the values for $m_2$ and $m_3$, and a range for $m_1$ 

\begin{eqnarray}
m_2 = (2.96 \pm 0.06) \times 10^{-2}eV,\nonumber\\
m_3 = (5.16 \pm 0.03) \times 10^{-2}eV, \nonumber \\
1.31 \times 10^{-2} eV < m_1 < 2.90 \times 10^{-2} eV. 
\end{eqnarray}
These masses are consistent with the latest upper bound on the neutrino mass from \cite{katrin}: $m_\nu < 0.45$ $eV$ (90\% CL).

\section{Neutrino oscillations from the Dirac equation}

In this section we review neutrino oscillations, initially focusing on the two flavor case, for subsequently focusing on the three flavor case. Consider the oscillations between the electron and the muon neutrinos. The Dirac equation for this case is

\begin{eqnarray}   \label{Diraceq}
i\gamma_\rho\partial^\rho\nu_e - \mu_e\nu_e -\zeta\nu_\mu = 0,\nonumber\\
i\gamma_\rho\partial^\rho\nu_e - \mu_\mu\nu_\mu -\zeta\nu_e = 0.
\end{eqnarray}
Here $\mu_\alpha$ and $\zeta$ correspond to the transition amplitudes between flavors, which are tied to the mass of the particle. We can express eq. (\ref{Diraceq}) as

\begin{eqnarray}
i\gamma_\rho\partial^\rho\nu_+ - m_+\nu_+ = 0,\nonumber\\ 
i\gamma_\rho\partial^\rho\nu_- - m_-\nu_- = 0,   
\end{eqnarray}
with 

\begin{equation}
m_{\pm} = \frac{\mu_\mu+\mu_e}{2}\pm\sqrt{\left(\frac{\mu_\mu-\mu_e}{2}\right)^2+\zeta^2}.    
\end{equation}
The mixing angle between these two flavors is

\begin{equation}
tan 2\theta=\frac{2\zeta}{\mu_\mu-\mu_e}.
\end{equation}
The details about the two-flavor oscillations, as well as the details about the derivation presented in this section, can be found in \cite{NewBook}.

\subsection{Three flavor oscillations}

Consider the three-flavor neutrino wave-function as 

%The mechanism that gives them mass is still not well understood.

\begin{equation}   \label{superposition}
\vert\psi_\nu(t)\rangle = C_e(t)\vert\nu_e\rangle + C_\mu(t)\vert\nu_\mu\rangle + C_\tau(t)\vert\nu_\tau\rangle.    
\end{equation}
\begin{figure}[h!]
\centering
\includegraphics[scale=0.4]{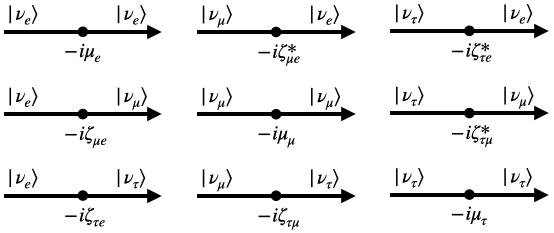} 
\caption{Transition amplitudes. The $\mu_\alpha$ are the trivial transition amplitudes, no flavor change, and the $\zeta_{\alpha\beta}$ are the transition amplitudes between flavors $\alpha$ and $\beta$. Adapted from \cite{NewBook}. \label{fig2}}
\end{figure}
The mixing matrix is the same as in eq. (\ref{matrix}), without CP violation. The wave-function in the mass basis is 
\begin{eqnarray}
\vert\psi_\nu(t)\rangle & = & D_1(0)e^{-iE_1 t}\vert\nu_1\rangle + \nonumber\\
& & D_2(0)e^{-iE_2 t}\vert\nu_2\rangle + \nonumber \\
& & D_3(0)e^{-iE_3 t}\vert\nu_3\rangle.     
\end{eqnarray}
This scenario assumes relativistic neutrinos for which the energies are approximately \cite{Exp}
\begin{equation}   \label{rel_neutrinos}
E_i = \sqrt{p^2_i + m^2_i} \approx p + \frac{m^2_i}{2E},    
\end{equation}
where $p$ and $E$ are the average momentum and energy of the neutrinos, respectively. Therefore, the transition probability to oscillate from a flavor $\alpha$ to a flavor $\beta$ is

\begin{equation}   \label{Probabilityof}
P_{\nu_\alpha\to\nu_\beta}=\sum_{i< j}\Lambda^{\alpha\beta}_{ij}e^{2i\Phi_{ij}},    
\end{equation}
with 

\begin{equation}   \label{Lamnda}
\Lambda^{\alpha\beta}_{ij}=U_{\alpha i}U^*_{\beta i}U^*_{\alpha j}U_{\beta j},
\end{equation}
and 

\begin{equation}
\Phi_{ij}=\frac{m_i^2-m_j^2}{4E}L,
\end{equation}
where $L$ is the length of the oscillation baseline. The Quantum Yang Baxter triangles presented in figure (\ref{fig1}), could be represented, in an equivalent way, considering the factors in eq. (\ref{Lamnda}). After the appropriate summation, we find that the transition probability from flavor $\alpha$ to flavor $\beta$ is the amplitude of the superposition of three different paths, as shown in figures (\ref{fig3}) and (\ref{fig4}).
\begin{figure}[h!]
\centering
\includegraphics[scale=0.42]{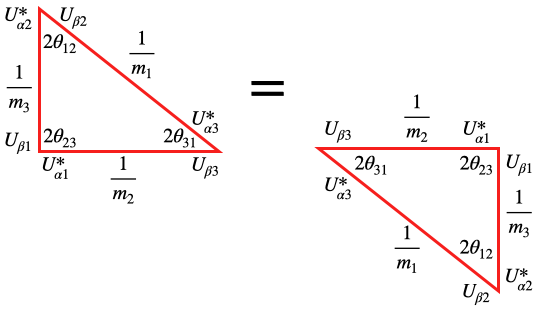} 
\caption{Inside the triangular relation, there are three different ways to evolve from a flavor $\alpha$ to a flavor $\beta$. \label{fig3}}
\end{figure}
\begin{figure}[h!]
\centering
\includegraphics[scale=0.4]{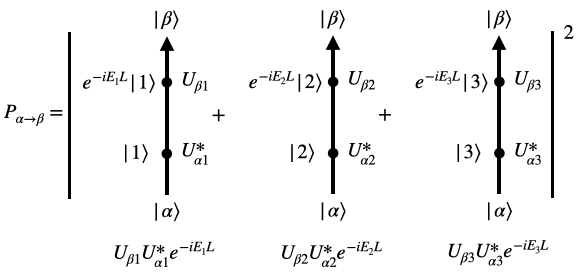} 
\caption{Transition probability from flavor $\alpha$ to flavor $\beta$. The three different ways to change from $\alpha$ to $\beta$. Adapted from \cite{NewBook}. \label{fig4}}
\end{figure}
The triangles in figure (\ref{fig3}) are consistent with the three different ways to change from flavor $\alpha$ to $\beta$ shown in figure (\ref{fig4}). However, figure (\ref{fig4}) is a more explicit illustration of the three possible paths from $\alpha$ to $\beta$, which interfere with each other. The three paths are related to the oscillation frequencies, which are connected to the three mass eigenvalues. \\
From eq. (\ref{Probabilityof}) it follows that the transition probability is
\begin{equation}   
P_{\nu_\alpha\to\nu_\beta} = \sum_{kl}\langle \nu_\beta\vert\nu_k\rangle \langle\nu_k\vert\nu_\alpha\rangle
\langle \nu_l\vert\nu_\beta\rangle \langle\nu_\alpha\vert\nu_l\rangle e^{-i(E_k-E_l)t}. \label{Prob12}
\end{equation}   
Therefore, the probability matrix is

\begin{eqnarray}   \label{ProbMatrix}
P^\beta_{\;\;\alpha} & = & \sum_{kln} \langle\nu_\beta\vert\nu_k\rangle \langle \nu_k\vert\nu_\alpha\rangle \times \nonumber \\
& & \langle \nu_l\vert\nu_\beta\rangle \langle \nu_\alpha\vert\nu_l \rangle \langle \nu_\beta\vert\nu_n \rangle \langle \nu_n\vert\nu_\alpha \rangle \times e^{-i(E_k-E_l)t} \nonumber \\
& = & R^{kl}_{\beta\alpha}R^{n\beta}_{\alpha l}R^{\beta\alpha}_{nk}e^{-i(E_k-E_l)t},
\end{eqnarray}
\begin{figure}[h!]
\centering
\includegraphics[scale=0.4]{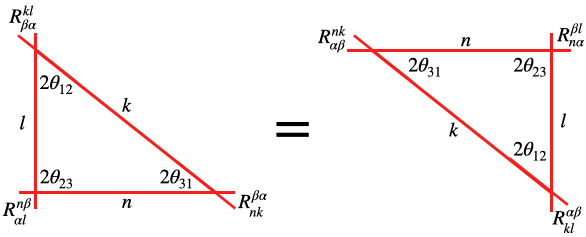} 
\caption{R-matrices corresponding to the QYBE. Each vertex of the triangle corresponds to one R-matrix. The internal lines $k$, $l$ and $n$ are summed over the three flavors inside the QYBE. Each R-matrix represents a possible path for the flavor transition $\alpha\to\beta$. \label{fig5}}
\end{figure}
where we introduced the completeness relation: $\hat{I}_{3\times3} = \sum\vert n\rangle \langle n\vert$. Eq. (\ref{ProbMatrix}) represents a probability matrix consistent with the QYBE, as $P^\beta_{\;\;\alpha} = \langle \alpha\vert P_{\nu_\alpha\to\nu_\beta}\vert\beta\rangle$. Eqns. (\ref{Prob12}) and (\ref{ProbMatrix}) give us a way to expand the triangles shown in figures (\ref{fig1}) and (\ref{fig3}) based on the graphical representation shown in figure (\ref{fig4}). If we represent the transition probability in eq. (\ref{Prob12}) using the QYBE, the triangles would correspond to those in figures (\ref{fig3}) and (\ref{fig4}). \\
The $R$-matrices in figure (\ref{fig5}) satisfy the QYBE
\begin{equation}   \label{R-matrix}
R^{kl}_{\beta\alpha}R^{n\beta}_{\alpha l}R^{\beta\alpha}_{nk}=R^{nk}_{\alpha\beta}R^{\beta l}_{n \alpha }R^{\alpha\beta}_{kl}, \\
\end{equation}
where
\begin{eqnarray}   \label{R-matrix2}
R^{kl}_{\beta\alpha} = \langle \nu_\beta\vert\nu_k\rangle \langle\nu_\alpha\vert\nu_l\rangle, \nonumber \\
R^{n\beta}_{\alpha l} = \langle \nu_\alpha\vert\nu_n\rangle \langle \nu_l\vert\nu_\beta\rangle, \nonumber \\
R^{\beta\alpha}_{nk} = \langle \nu_n\vert\nu_\beta\rangle \langle \nu_k\vert\nu_\alpha\rangle,
\end{eqnarray}
for the left-hand side, and akin results for the right-hand side. The R-matrices that appear in front each vertex of the triangles in figure (\ref{fig4}) could be interpreted as transition matrices showing the different paths for the change of flavor $\alpha\to\beta$. \\

%In this way, we showed that the pattern of flavor oscillation for neutrinos is completely consistent with the QYBE.
Eqns. (\ref{Prob12}) and (\ref{ProbMatrix}) anticipate a hidden symmetry which is not spontaneously broken, as shown in \cite{M1, M2, M3}. The conditions for spontaneously breaking a symmetry are well known, and are summarized in \cite{zeromass}.

%The justification for this additional term comes from the fact that we can express the probability (\ref{Prob12}) as the sum of three terms as follows

%\begin{eqnarray}   \label{Prob1-3}
%P_{\nu_\alpha\to\nu_\beta}=\frac{1}{3}\sum_{kl}<\nu_\beta\vert\nu_k><\nu_k\vert\nu_\alpha>\times\nonumber\\<\nu_l\vert\nu_\beta><\nu_\alpha\vert\nu_l>e^{-i\frac{(m_k-m_l)}{\gamma}t}+\nonumber\\
%\frac{1}{3}\sum_{kn}<\nu_\beta\vert\nu_k><\nu_k\vert\nu_\alpha>\times\nonumber\\
%<\nu_n\vert\nu_\beta><\nu_\alpha\vert\nu_n>e^{-i\frac{(m_k-m_n)}{\gamma}t}+\nonumber\\
%\frac{1}{3}\sum_{ln}<\nu_\beta\vert\nu_n><\nu_n\vert\nu_\alpha>\times\nonumber\\<\nu_l\vert\nu_\beta><\nu_\alpha\vert\nu_l>e^{-i\frac{(m_n-m_l)}{\gamma}t}.
%\end{eqnarray}
%
%This way of expressing the probability of transition between the flavor $\alpha$ towards the flavor $\beta$, is a more explicit way to show that there are three possible trajectories for the neutrino to have the mentioned flavor change. The expression (\ref{Prob12}) is equivalent to the expression (\ref{Prob1-3}). 

\section{Conclusions}   \label{concl}

In this paper we have shown that neutrino oscillations are consistent with the superposition of three different histories for the flavor transition $\alpha\to\beta$. This is also consistent with the QYBE, which constraint the relationship between the mixing angles and the mass eigenvalues. These constraints helped us to explicitly calculate the mass eigenvalue $m_1$, $m_2$ and $m_3$. In the QYBE framework, each side of the triangle corresponds to each history of the flavor transition $\alpha\to\beta$. Within the same scenario, we defined a probability matrix $P^\beta_{\;\;\alpha}$, where each element corresponds to a transition probability between flavors. The matrix is equivalent to the QYBE. Our formulation suggests that the flavor oscillations in the neutrino sector are related to a hidden symmetry, which is not spontaneously broken. Instead, neutrinos oscillate between three different vacuum states, corresponding to the different flavor states, in a recurrent way, without selecting a particular ground state. Nevertheless, further studies are necessary to understand the underlying mechanism that generate neutrino masses.

% The \nocite command causes all entries in a bibliography to be printed out
% whether or not they are actually referenced in the text. This is appropriate
% for the sample file to show the different styles of references, but authors
% most likely will not want to use it.
\nocite{*}

%\bibliography{apssamp}% Produces the bibliography via BibTeX.

\end{document}